

\magnification=1200
\baselineskip=16pt
\hskip 10 cm
CIEA-GR-9407

\hskip 10cm
ICN-UNAM-9405

\hskip 10cm

\vskip2pc
\centerline{\bf GEOMETRY OF DEFORMATIONS OF RELATIVISTIC MEMBRANES}
\vskip2pc
\centerline {\bf Riccardo Capovilla$^{(1)}$ and Jemal Guven$^{(2)}$}
\vskip.5pc
\centerline {\it (Revised Version)}
\vskip.5pc
\it
\centerline {$^{(1)}$Departamento de F\'{\i}sica}
\centerline{  Centro de Investigaci\'on y de Estudios Avanzados del
I.P.N. }
\centerline {Apdo. Postal 14-740, 07000 M\'exico, D. F., MEXICO} \rm
\centerline{(capo@fis.cinvestav.mx)}
\vskip1pc
\it
\centerline {$^{(2)}$ Instituto de Ciencias Nucleares}
\centerline {Universidad Nacional Aut\'onoma de M\'exico}
\centerline {Apdo. Postal 70-543, 04510 M\'exico, D. F., MEXICO} \rm
\centerline{(guven@roxanne.nuclecu.unam.mx)}
\vskip1pc

\centerline{\bf Abstract}
\vskip1pc
{\leftskip=1.5cm\rightskip=1.5cm\smallskip\noindent
A kinematical description of infinitesimal deformations of the
worldsheet spanned in spacetime by a relativistic
membrane is presented. This provides a framework for
obtaining both the classical equations of motion and the
equations describing infinitesimal deformations about solutions
of these equations when the action describing the
dynamics of this membrane is constructed using {\it any}
local geometrical worldsheet scalars. As examples, we consider a
Nambu membrane, and an action quadratic in the extrinsic curvature of
the worldsheet.
 \smallskip}
\vfill

\vfill\eject
\noindent{\bf I. INTRODUCTION}
\vskip1pc

A useful approximate description of the dynamics of many physical
systems is to model them as relativistic membranes of an
appropriate dimension. The construction of the
corresponding phenomenological action
determining the dynamics of the membrane involves the
selection of an appropriate linear combination of the geometrical scalars
of its worldsheet.
At lowest order, this action is proportional to the intrinsic volume of the
worldsheet, and it has come to be known as the Nambu action.
If the approximation stops here, the classical trajectory of
the membrane will be an extremal surface of the background spacetime.
A large body of information has accumulated
on the dynamics of geometrically symmetrical extremal solutions
(see {\it e.g.} Ref. [1], for a review in the context of cosmology).
To place these solutions in proper context, however, their stability needs to
be examined both with respect to classical and quantum mechanical perturbations
propagating on
What becomes clear when this is attempted for even the simplest models
is that a manifestly covariant formalism to describe the evolution of these
perturbations  which is also independent of the
particular symmetry of the membrane is desirable.
This problem was approached for Nambu membranes by one of the
authors in Ref.[3] and, independently, using similar
techniques in Refs.[4] and [5].\footnote * {The relevant mathematical
formalism was developed by mathematicians much earlier
in the context of minimal surfaces [6].}
The perturbation is described by
a system of coupled linear wave equations, one for
the projection of the infinitesimal deformation in the worldsheet onto
each normal direction, which can be considered as scalar fields
living on the worldsheet. In this way, a perturbative framework for
examining the stability of any system described by the Nambu action
is provided.

The analysis presented in Refs. [3,4] and [5]
was tailored to describe extremal surfaces. For some time, however,
it has been realized that extrinsic curvature additions
or corrections to the Nambu action can have a dramatic
influence on the dynamics on short length scales [7,8].
In particular, when such corrections are introduced
the development of cusps or kinks on the membrane appears to be
inhibited on these scales. These corrections may arise in a more realistic
truncation of the underlying field theory,
or may be induced by quantum mechanical fluctuations.

The equations of motion which correspond to a generic action
which is quadratic in the extrinsic curvature, are typically
hyperbolic equations which are fourth order in derivatives
of the embedding functions describing the worldsheet.
To derive these equations and their subsequent linearizations,
one could attempt to imitate the analysis applied earlier
to extremal surfaces. However, by following
this case by case approach, one can easily lose sight of the fact
that there is a solid kinematical structure underpinning
these equations which is entirely independent of the
underlying dynamics. In this paper,
we develop such a kinematical framework for describing
deformations of an arbitrary worldsheet. Relevant earlier
investigations in this direction are Refs.[5,8,9] and [10].
In Ref.[10] Hartley and Tucker exploit very elegant
exterior differential techniques to derive the equations of motion
for relativistic membranes. This language could, potentially,
provide a very powerful geometrical approach to the description of
deformations.

Our effort divides naturally into an examination of the
deformation of the intrinsic and of the extrinsic geometry of the
worldsheet.  Once this is done, both the equations of motion,
and the equations describing the dynamics of deformations
about classical solutions can be constructed in Lego block fashion,
by assembling the various kinematical ingredients.

The paper is organized as follows.
To establish our notation, we begin in section II by summarizing
the well known classical kinematical description provided by the
Gauss-Weingarten equations of an embedded timelike worldsheet of dimension
$D$ in a spacetime of dimension $N$, in terms of its
intrinsic and extrinsic geometry [11,12,13]. There are two
structures which describe the extrinsic geometry. One of these is
given by the extrinsic curvatures, and is well understood. The other
structure, which we  call the extrinsic twist potential, only
features when the co-dimension of the worldsheet is two or higher.
The extrinsic twist plays a subtle role related to
the covariance of the description of the geometry under rotations of the
normals to the worldsheet.

In this same kinematical spirit, in
sections III and IV we describe the deformation
of the worldsheet. There are analogues of the Gauss-Weingarten equations
which are useful for identifying the structures associated with
such deformations. The description of the deformation divides
naturally into two parts. The deformation of the intrinsic geometry
is very simple to describe. Indeed, for example, the deformations
of the worldsheet metric  provide
a geometrically satisfying definition of the extrinsic curvatures.
The description of the deformation of the extrinsic
geometry is less simple. One reason for this is because the
naive deformation of the structures associated with the
extrinsic geometry do not transform covariantly
under normal frame rotations. By examining the
deformation of the normal vectors (the analogue of the
Weingarten equations) we can
identify a `connection' which guarantees the manifest covariance
of the deformation of such structures under normal frame rotations.
It turns out, however, that this connection does not appear
in any physical quantity, and thus it does not need to be
calculated.

In section V, we apply this  kinematical framework to some phenomenological
actions  of physical interest. We consider first the familiar Nambu action,
to show how our analysis reproduces the results of  Ref. [3] and [4].
Next, we consider an action quadratic in the extrinsic curvature. We derive
the equations of motion, and the equations that describe the
dynamics of deformations about classical solutions in the case the
background spacetime is Minkowski space. We conclude in section VI with a
brief discussion.

We confine our attention to closed membranes without physical boundaries.

\vskip2pc
\noindent{\bf II. MATHEMATICS OF THE EMBEDDING OF THE WORLDSHEET}
\vskip1pc

In this section we provide an overview of the well known mathematical
description of the worldsheet of a membrane viewed as an embedded
surface in a fixed background spacetime [11,12,13].

Let us consider an oriented {\it timelike} worldsheet $m$ of dimension $D$
described by the embedding

$$x^\mu=X^\mu(\xi^a)\,,\eqno(2.1)$$
$\mu=0,\cdots, N-1$, and $a=0,\cdots, D-1$,
in an $N$-dimensional spacetime $M$,
endowed with the metric $g_{\mu\nu}$. The $D$ vectors

$$e_a :=X^\mu_{,a}\partial_\mu\eqno(2.2)$$
form a basis of tangent vectors to $m$ at each point of $m$.
The metric induced on the world sheet is then
given by

$$\gamma_{ab}= X^\mu_{,a}
X^\nu_{,b}\,g_{\mu\nu} = g(e_a,e_b)\,.\eqno(2.3)$$
The signature of $\gamma_{ab}$ is taken to be $\{-,+,...,+ \}$.

Let $n^{i}$ denote the $i^{\rm th}$ unit normal to the worldsheet,
$i=1,\cdots, N-D$, defined by

$$ g(n^i,n^j) = \delta^{ij}, \; \; g(e_a,n^i) = 0\,. \eqno(2.4)$$
It is important to emphasize that these equations define
the normal vector fields $n^i$ only up to a $O(N-D)$ rotation, and  up to
a sign.

Normal vielbein indices are raised and
lowered with $\delta^{ij}$ and $\delta_{ij}$, respectively, whereas
tangential indices are raised and lowered with $\gamma^{ab}$ and
$\gamma_{ab}$, respectively.

The collection of vectors
$\{e_a, n^i\}$ can be used as a basis for the spacetime vectors
appropriate for the geometry under consideration.

We define the worldsheet projections of the
spacetime covariant derivatives with $D_a := e_a^\mu D_\mu$,
where $D_\mu$ is the (torsionless) covariant derivative compatible
with $g_{\mu \nu}$. Let us now consider the
worldsheet gradients of the basis vectors
$\{e_a, n^i\}$, $D_a e_b$ and $D_a n^i$.
These spacetime vectors  can always
be decomposed with respect to the basis,\footnote *
{To avoid confusion, we adopt the notation $\omega$
(adopted by Maeda and Turok in Ref.[6]) instead
of $T$ as was used in Ref.[3])
for the twist potential, and $\Omega$ (adopted by
Carter in Ref.[5]) for the corresponding curvature.}

$$\eqalignno{D_a e_b =& \gamma_{ab}{}^c e_c - K_{ab}{}^i n_i\,,  &(2.5a)\cr
	   D_a n^i =& K_{ab}{}^i e^b + \omega_a{}^{ij} n_j\,.&(2.5b)\cr}$$
These kinematical expressions,
generalizing the classical Gauss-Weingarten equations,
describe completely the extrinsic
geometry of the worldsheet.

The $\gamma_{ab}{}^c$ are the
connection coefficients compatible with the worldsheet metric
$\gamma_{ab}$,

$$\gamma_{ab}{}^c =g(D_a e_b, e^c)=\gamma_{ba}{}^c\,.\eqno(2.6)$$
The quantity
$K_{ab}{}^i$ is the $i^{\rm th}$ extrinsic curvature of the worldsheet,

$$K_{ab}{}^i=-g(D_a e_b,n^i) =K_{ba}{}^i\,.\eqno(2.7)$$
The symmetry in the tangential indices of these quantities is
a consequence of the integrability of the tangential basis $\{e_a\}$.

The normal fundamental form, or extrinsic twist potential,
of the worldsheet is defined by

$$\omega_a{}^{ij} = g(D_a n^i,n^j) =-\omega_a{}^{ji}\,.\eqno(2.8)$$
In the familiar case of a hypersurface embedding,
$D=N-1$, the extrinsic twist vanishes identically.
The geometrical meaning of $\omega_a{}^{ij}$
can be understood by recalling that
there is the freedom to perform local rotations
 of the normal frame, $\{ n^i\}$. With respect to the rotation,
$n^i\to O^{i}{}_{j} n^j$, $\gamma_{ab}{}^c$ transforms as
a scalar, and $K_{ab}{}^i$ transforms as a vector. The extrinsic
twist potential,
$\omega_a{}^{ij}$, transforms as a connection,

$$\omega_a\to O \omega_a O^{-1} + O_{,a} O^{-1}\,. \eqno(2.9)$$
As discussed in detail in Ref.[3], for example,
it can be considered as the gauge field
associated with the normal frame rotation group.
It is desirable to implement this covariance
in a manifest way. Let 	$\nabla_a$ be the (torsionless)
covariant derivative compatible with $\gamma_{ab}$ induced
on $m$. We introduce a new worldsheet covariant derivative
$\tilde\nabla_a$ defined on fields transforming as tensors under
normal frame rotations
as follows,

$$\tilde\nabla_a \Phi^i{}_j :=
\nabla_a \Phi^i{}_j - \omega_a{}^{ik}\Phi_{kj}
- \omega_{a\,jk}\Phi^{ik}\,.\eqno(2.10)$$
We also introduce  the curvature
associated with $\omega_a{}^{ij}$ with

$$\Omega_{ab}{}^{ij}:=\nabla_b \omega_a{}^{ij} - \nabla_a \omega_b{}^{ij} -
 \omega_b{}^{ik} \omega_{a\,k}{}^{j} +  \omega_a{}^{ik}
\omega_{b\,k}{}^{j}\,.\eqno(2.11)$$
Note that when $D=1$, $\Omega_{ab}{}^{ij}=0$ so that
$\omega_a{}^{ij}$ is pure gauge, at least locally.
We also know that when $D=N-1$, $\omega_a{}^{ij}=0$. When $D=N-2$,
the gauge group is $O(2)$ with a single generator. We conclude that
the non-abelian (or non-linear) character of
$\omega_a{}^{ij}$ displayed in an arbitrary dimension
does not manifest itself in spacetime dimensions lower than five.

Given some specification of intrinsic and
extrinsic geometry Eqs.(2.5) will not
generally be consistent with any embedding because $X^\mu(\xi)$ is
over specified by these equations.
Consistency will require that the intrinsic
and extrinsic geometry satisfy the
Gauss-Codazzi, Codazzi-Mainardi, and Ricci integrability conditions,

$$g(R(e_b,e_a)e_c,e_d) = {\cal R}_{abcd} -
K_{ac}{}^{i} K_{bd\,i} + K_{ad}{}^i K_{bc\,i}
\,,\eqno(2.12a)$$

$$g(R(e_b,e_a)e_c, n^{i}) = \tilde\nabla_a K_{bc}{}^{i} -
\tilde\nabla_b K_{ac}{}^{i}\,,\eqno(2.12b)$$

$$g(R(e_b,e_a)n^{i},n^{j}) =
 \Omega_{ab}{}^{ij} -
K_{ac}{}^{i} K_b{}^{c\,j}
+ K_{bc}{}^i K_a{}^{c\,j}\,.\eqno(2.12c)$$
We use the notation
$g(R(Y_1, Y_2)Y_3,Y_4) =
 R_{\alpha\beta\mu\nu}Y_1^\mu Y_2^\nu Y_3^\beta Y_4^\alpha$.
$R^\alpha{}_{\beta\mu\nu}$ is the Riemann tensor of the
spacetime covariant derivative $D_\mu$, whereas ${\cal R}^a{}_{bcd}$ is
the Riemann tensor of the worldsheet covariant derivative $\nabla_a$.
Note that Eq.(2.12c) possesses no non-trivial contractions.
In particular, it is vacuous when $D=1$, and when $D=N-1$.

These equations can be obtained directly from the Gauss-Weingarten equations,
by taking a second spacetime covariant derivative
projected onto the worldsheet, subtracting the same equation with
the indices on the derivatives switched, and
exploiting the spacetime Ricci identity.

In de Sitter spacetime,

$$R_{\mu\nu\alpha\beta}= H^2(g_{\mu\alpha} g_{\nu\beta}
- g_{\mu\beta} g_{\nu\alpha})\,,\eqno(2.13)$$
so that

$$g(R(e_b,e_a)e_c,e_d) =H^2 (\gamma_{ac}\gamma_{bd}
-\gamma_{ad}\gamma_{bc})\,.\eqno(2.14)$$
The right hand sides of Eqs.(2.12b) and (2.12c)
both vanish.\footnote *
{From Eq. (2.12a), it is clear that the necessary and sufficient
condition that the worldsheet will also be a de Sitter space is that
$$K_{ac}{}^{i} K_{bd\,i} - K_{ad}{}^i K_{bc\,i}
\propto (\gamma_{ac}\gamma_{bd}-\gamma_{ad}\gamma_{bc})\,.$$}

The appearance of the curvature of the
extrinsic twist potential, $\Omega_{ab}{}^{ij}$, in the Ricci
integrability condition, Eq.($2.12c$),
provides us with additional information about
the extrinsic twist itself.
First, note that, for a given spacetime, Eq. ($2.12c$) implies that
the curvature $\Omega_{ab}{}^{ij}$ is completely determined once the
intrinsic geometry and the extrinsic curvatures
are specified. In fact, as was emphasized by Carter, $\Omega_{ab}{}^{ij}$
is the conformally invariant tracefree part of the squared extrinsic
curvature [13,14]. This equation also provides the necessary and
sufficient conditions that the extrinsic twist can be gauged away,

$$g(R(e_a,e_b)n^{i},n^{j}) -
K_{ac}{}^{i} K_b{}^{c\,j}
+ K_{bc}{}^i K_a{}^{c\,j} = 0\,.\eqno(2.15)$$
In particular, in de Sitter spacetime,
if all but one $K^i_{ab}$ vanish, then the antisymmetric
product of extrinsic curvature tensors vanishes, and the
integrability condition is satisfied automatically.

\vskip2pc
\noindent{\bf III. DEFORMATIONS OF THE INTRINSIC GEOMETRY}
\vskip1pc

In the previous section, we have described
the characterization of a single
embedded surface in spacetime, in terms of
its intrinsic and extrinsic geometry.

Let us now consider the neighboring surface described by a
deformation of $m$,

$$x^\mu=X^\mu(\xi^a)+\delta X^\mu(\xi^a)\,.$$
We can decompose the infinitesimal deformation
vector field $\delta X^\mu$ with respect to the spacetime
basis $\{ e_a ,n^i \}$, as,

$$\delta X = \Phi^a e_a  + \Phi^i n_i \,.\eqno(3.1)$$
The tangential projection can always be identified with the
action of a worldsheet diffeomorphism, $\delta X^\mu
= \Phi^a X^\mu_a$,
and so will subsequently be ignored. The physically observable
measure of the deformation is therefore provided by the projection of
$\delta X^\mu$ orthogonal to $m$, characterized by the
$N-D$  scalar fields $\Phi^i$.

Our task will be to express the deformation of the geometrical
structures introduced in section II as linear combinations of the
scalar fields $\Phi^i$, and their covariant derivatives,
$\tilde\nabla_a \Phi^i$,
$\tilde\nabla_a\tilde\nabla_b \Phi^i$, $\cdots$ We make use of
 the
covariant derivative defined in (2.10), because $\Phi^i$ transforms
as a vector under normal frame rotations.

In this section we consider the deformation of the {\it intrinsic}
geometry of the worldsheet under a deformation in the embedding.
The displacement $\delta X^\mu$ in the embedding induces
a displacement in the tangent basis $\{e_a\}$. In light of the discussion
above, let $\delta  =\Phi^i n_i$, and consider the gradients of $\{e_a\}$
along the vector field $\delta $, defined with $D_{\delta } :=
\delta^\mu D_\mu$. We can always expand
$D_{\delta } e_a$ with respect to the spacetime basis
$\{e_a,n_i\}$, in a way
analogous to the Gauss equation, (2.4a), as follows,

$$D_{\delta } e_a = \beta_{ab} e^b + J_{a\,j} n^j\,.\eqno(3.2)$$
Comparison with the Gauss equation shows that
the quantity $\beta_{ab}$, defined by

$$\beta_{ab}=g(D_{\delta }e_a, e_b)=\beta_{ba}\,,\eqno(3.3)$$
appears in the same position as $\gamma_{ab}{}^c$.
The quantities $J_{a\,i}$ are defined by

$$J_{a\,i} = g(D_{\delta }e_a, n_i)\,,\eqno(3.4)$$
and appear in the same position as $K_{ab}{}^i$ in
the Gauss equation.
We note that $\beta_{ab}$ transforms as a scalar under normal frame
rotations, whereas
$J_{a\,i}$ transforms as a vector.

In order to express
$\beta_{ab}$ and $J_{a\,i}$ in terms of $\Phi^i$ and its covariant derivatives
it is crucial to recognize that, for all infinitesimal deformations
of the worldsheet [15],
one has

$$D_{\delta } e_a = D_a \delta\,.\eqno(3.5)$$
In words, this equation follows from the equality of
the gradient along the deformation vector field $\delta$ of
the tangential basis $\{ e^a \}$, with the change of $\{ e^a \}$
induced by the displacement of the worldsheet.

Using Eq. (3.5), it is easy to show that,

$$\beta_{ab}=g(D_{\delta } e_a, e_b) = g(D_a\delta , e_b)
= g(D_a n^i, e_b)\Phi_i = K_{ab}{}^i \Phi_i \,, \eqno(3.6)$$

$$J_{a\,i} = g(D_{\delta } e_a, n_i)
= g(D_a\delta , n_i)
= g( D_a n^j, n^i)\Phi_j + \nabla_a\Phi_i
=\tilde \nabla_a\Phi_i\,.\eqno(3.7)$$
Therefore, the gradients along the deformation of the tangential
vectors depend on the values of the scalar fields $\Phi^i$,
and on their first derivatives along the worldsheet.

The deformation in the induced metric on $m$ is
just twice $\beta_{ab}$,

$$D_\delta \gamma_{ab} = D_\delta g(e_a,e_b)= 2 g(e_a, D_\delta e_b)
=2 \beta_{ab} = 2 K_{ab}{}^i \Phi_i\,.\eqno(3.8)$$
In fact, this equation encodes the geometrical role of
$K_{ab}^i$. It is half
the change induced in the worldsheet metric per unit
proper deformation of the worldsheet along the $i^{\rm th}$ normal direction.

This is  all we need to know about the deformation of
the intrinsic geometry, if we are only interested in the
deformation of extremal surfaces.
However, one might also be interested in more general theories
that contain scalars
constructed with the worldsheet curvature tensor,
${\cal R}^a{}_{bcd}$.

To derive an expression for the deformation of
${\cal R}^a{}_{bcd}$,  we exploit the Palatini identity, and
Eq.(3.8), to write the tensor valued infinitesimal
deformation of the worldsheet Christoffel symbol,

$$\eqalign{D_{\delta }\gamma_{ab}{}^c &={1\over 2}\gamma^{cd}\left[
\nabla_b(D_{\delta }\gamma_{ad}) + \nabla_a (D_{\delta }
\gamma_{bd}) -
\nabla_d (D_{\delta }\gamma_{ab})\right]\cr
&= \gamma^{cd}\left[
\nabla_b(K_{ad}{}^i\Phi_i) + \nabla_a (K_{bd}{}^i\Phi_i) -
\nabla_d (K_{ab}{}^i\Phi_i)\right]\,.\cr}\eqno(3.9)$$
The infinitesimal deformation in the
worldsheet Riemann tensor then can be simply expressed
in terms of worldsheet covariant derivatives of the
$D_{\delta }\gamma_{ab}{}^c$,

$$ D_{\delta } {\cal R}^a{}_{bcd} =
\nabla_c( D_{\delta }\gamma_{bd}{}^a) -
\nabla_d(  D_{\delta } \gamma_{bc}{}^a)\,. \eqno(3.10)$$
We see that it depends on second and first worldsheet
derivatives of the
scalar fields $\Phi^i$.

The corresponding infinitesimal variations in
the Ricci tensor and the scalar curvature are, respectively,

$$ D_{\delta } {\cal R}_{ab}  =
\nabla_c ( D_{\delta } \gamma_{ab}{}^c) -
\nabla_b( D_{\delta } \gamma_{ac}{}^c)\,, $$

$$ D_{\delta }{\cal R}=
 \nabla^a( \gamma^{bc} D_{\delta } \gamma_{bc}{}^a) -
\nabla^b ( D_{\delta }\gamma_{ab}{}^a)
- 2{\cal R}_{ab} K^{ab}{}_i\Phi^i \,.$$
Thus, modulo a divergence,

$$D_{\delta }{\cal R}= - 2{\cal R}_{ab} K^{ab}{}_i\Phi^i \,,\eqno(3.11)$$
We also note that,

$$D_{\delta } (\sqrt{-\gamma}{\cal R}) =
- 2{\cal G}_{ab} K^{ab}{}_i\Phi^i \,, \eqno(3.12)$$
where ${\cal G}_{ab}$ is the worldsheet Einstein tensor.

This concludes the analysis of the deformation of the intrinsic
geometry of the worldsheet.

\vskip2pc
\noindent{\bf IV. DEFORMATIONS OF THE EXTRINSIC GEOMETRY}
\vskip1pc

The extrinsic geometry is characterized by the extrinsic
curvatures, $K_{ab}{}^i$, and the extrinsic twist, $\omega_a{}^{ij}$.
As a preliminary step, let us examine the gradient along the
deformation vector field of the normal basis, $D_{\delta } n^i$,
in the same way as we did for the tangent basis. We expand,

$$ D_{\delta } n_i =  - J_{a\,i} e^a + \gamma_{ij} n^j\,.\eqno(4.1)$$
This equation is the analogue for infinitesimal
deformations of the Weingarten equation (2.4b).
We note that $J_{a\,i}$ appears in Eqs.(3.2) and (4.1)
in an analogous way to that of $K_{ab}{}^i$ in Eqs.(2.4a) and (2.4b).

The normal projection of $D_{\delta }n_i$,

$$\gamma_{ij} = g(D_{\delta } n_i , n_j ) = -\gamma_{ji}\,,\eqno(4.2)$$
is a new structure we have not encountered already.
It vanishes on a hypersurface embedding, in the same way that
$\omega_a{}^{ij}$ vanishes in the corresponding Weingarten equation.
In contrast to $J_{a\,i}$ and $\beta_{ab}$, however, there is no
simple relationship between $\gamma_{ij}$ and deformations
of the worldsheet analogous to Eqs.(3.6) and (3.7).

The analogy between Eq. (4.1) and the
Weingarten equation suggests a role
for $\gamma_{ij}$ analogous to $\omega_a{}^{ij}$. In
particular, $\gamma_{ij}$, like $\omega_a{}^{ij}$,
transforms as a connection under normal frame
rotations

$$\gamma \to O \gamma O^{-1} + (D_{\delta } O )\, O^{-1}\,.\eqno(4.3)$$
However, by an appropriate choice of $D_{\delta }O$, it
is always possible to gauge $\gamma_{ij}$ away on the worldsheet.
Reflecting this fact, as we will demonstrate below,
$\gamma_{ij}$  will never appear explicitly in any physical
quantity, although it will show up in
intermediate calculations.
Nonetheless, we will insist
on explicit covariance under normal frame rotations.
For this purpose, we introduce
a covariant deformation derivative as follows,

$$\tilde D_\delta \Psi_i = D_\delta \Psi_i -\gamma_{i}{}^j \Psi_j\,.
\eqno(4.4)$$
Eq. (4.1) can then be written in the form

$$
\tilde D_\delta n_i = - J_{ai} e^a = - (\tilde \nabla_a \Phi_i) e^a \,.
\eqno(4.5)
$$
\vfill\eject
\noindent{\bf IV.1 Deformations of the extrinsic curvature}
\vskip1pc

Let us now evaluate  the deformation
of the extrinsic curvatures, $\tilde D_\delta K_{ab}{} ^i$.
Using its definition we have that,

$$ \tilde D_{\delta } K_{ab}{}^i = - g (\tilde D_{\delta } n^i,
 D_a e_b )
- g (n^i , D_{\delta } D_a e_b )\,. $$
Using Eq. (4.5), and the Gauss equation ($2.5a$),
 the first term on the right hand side is given by

$$
- g (\tilde D_{\delta } n^i, D_a e_b ) =  \gamma_{ab}{}^c J_c{}^i
\,.
$$
The second term on the right hand side  can be developed
using the Ricci identity, as,

$$\eqalign{ - g (n^i , D_{\delta } D_a e_b ) =&
- g (n^i, R (\delta , e_a ) e_b ) - g (n^i, D_a D_{\delta } e_b ) \cr
=& -  g (n^i, R ( n_j , e_a ) e_b ) \Phi^j - D_a g (n^i, D_\delta e_b )
+ g (D_a n^i  ,  D_\delta e_b  ) \cr
=&  - g (n^i, R ( n_j , e_a ) e_b ) \Phi^j - D_a J_b{}^i
+ \beta_{bc} K_a{}^{c\,i}
+ \omega_a{}^{ij} J_{bj} \cr
=&  - g (n^i, R ( n_j , e_a ) e_b ) \Phi^j - \tilde\nabla_a\tilde\nabla_b
\Phi^i + K_{bc j} K_a{}^{c\, i} \Phi^j\,, \cr }  $$
where in the last line we have used Eqs. (3.6) and (3.7).

Therefore  we find

$$\tilde D_{\delta } K_{ab}{}^i=
-\tilde\nabla_a\tilde\nabla_b \Phi^i +
\left[g(R(e_a, n_j )e_b, n^i)
 + K_{ac}{}^i K^c{}_{b\,j}\right]\Phi^j\,. \eqno(4.6)$$
Note that the change of the extrinsic curvatures under an infinitesimal
deformation of the worldsheet
involves second derivatives of the scalar fields $\Phi^i$.

The left hand side of Eq. (4.6) is manifestly symmetric in the
indices $a$ and
$b$. The apparent integrability condition on the right hand side,

$$ 2 \tilde\nabla_{[a}\tilde\nabla_{b]} \Phi^i =
\left[g(R(e_a, n_j)e_b,n^i)
 + K_{ac}{}^i K^c{}_{b\,j} -(a\leftrightarrow b) \right]\Phi^j \,,
\eqno(4.7)$$
is automatically  satisfied as a consequence of the
Ricci integrability condition ($2.12c$). To show this, one needs
to use the cyclic Bianchi identities for the spacetime
Riemann tensor, $R_{\alpha [\beta \mu \nu ] } = 0 $, and the identity
$2 \tilde\nabla_{[a}\tilde\nabla_{b]} \Phi^i = \Omega_{ab}{}^i{}_j \Phi^j$.
\vfill\eject
\noindent{\bf IV.2 Deformations of the extrinsic twist potential}
\vskip1pc

We turn now to the analysis of the deformation of the
estrinsic twist $\omega_a{}^{ij}$.
Unfortunately, the obvious measure of the deformation,
$\tilde D_\delta \omega_a{}^{ij}$, does not transform
covariantly under normal frame rotations.
However, by examining $\tilde D_\delta \omega_a{}^{ij}$ itself, we can identify
the appropriate addition that provides a covariant measure of the
deformation.

By definition, we have that

$$\tilde D_\delta \omega_a{}^{ij} =
D_\delta \omega_a{}^{ij} -\gamma^i{}_k \omega_a{}^{kj} -
\gamma^j{}_k \omega_a{}^{ik}\,, \eqno(4.8) $$
where

$$D_\delta \omega_a{}^{ij} = D_\delta g(D_a n^i, n^j)
=  g( D_a n^i, D_\delta n^j) +
g(D_\delta D_a n^i, n^j)\,.\eqno(4.9) $$
The first term on the right hand side is

$$ \eqalignno{ g( D_a n^i, D_\delta n^j) =&
K_{ab}{}^i g( e^b, D_\delta n^j) + \omega_a{}^{ik} g (n_k , D_\delta n^j)
\cr
=& - K_{ab}{}^i J^{bj} + \omega_a{}^{ik} \gamma^j{}_k \cr
=& - K_{ab}{}^i \tilde\nabla^b \Phi^j  + \omega_a{}^{ik}
\gamma^j{}_k\,.&(4.10)\cr
}
$$
In the second term of (4.9), using the Ricci identity, we have,

$$ \eqalignno{ g(D_\delta D_a n^i, n^j) =&  g( R(\delta, e_a) n^i, n^j)
+ g( D_a D_\delta n^i , n^j ) \cr
=&  g( R(\delta, e_a) n^i, n^j) + D_a g(D_\delta n^i , n^j )
- g (D_\delta n^i , D_a n^j) \cr
=&  g( R(\delta, e_a) n^i, n^j) + \nabla_a \gamma^{ij}
+ K_{ab}{}^j \tilde\nabla^b \Phi^i  - \omega_a{}^{jk} \gamma^i{}_k\,.
&(4.11)\cr
}$$
where we have used (4.10) in the last line.

We find then that

$$\tilde D_\delta \omega_a{}^{ij} = - K_{ab}{}^i \tilde\nabla^b \Phi^j +
K_{ab}{}^j \tilde\nabla^b \Phi^i + \nabla_a \gamma^{ij} +
g(R(n_k, e_a) n^j, n^i)\Phi^k\,, \eqno(4.12a)$$
or

$$\tilde D_\delta \omega_a{}^{ij}  - \nabla_a \gamma^{ij}
= - K_{ab}{}^i \tilde\nabla^b \Phi^j +
K_{ab}{}^j \tilde\nabla^b \Phi^i  +
g(R(n_k, e_a) n^j, n^i)\Phi^k\,. \eqno(4.12b)$$
This result indicates that the left hand side
is the covariant measure of the
deformation of the extrinsic twist potential.  In fact,
the right hand side of  $(4.12b)$ is manifestly covariant,
and  thus so also is the left hand side.
Both sides of Eq.($4.12b$) are manifestly antisymmetric in
the indices $i$ and $j$. Unlike for the deformation of
the extrinsic curvature, Eq.(4.6), here
no integrability condition need ever be invoked.
We also note the identity

$$\tilde D_\delta \omega_a{}^{ij} - \nabla_a \gamma^{ij} =
D_\delta \omega_a{}^{ij} -
\tilde\nabla_a \gamma^{ij} \,.\eqno(4.13)$$

The deformation of the curvature of the
extrinsic twist is then given by

$$ D_{\delta } \Omega_{ab}{}^{ij} = \tilde\nabla_a  (D_\delta
\omega_b{}^{ij}) - \tilde\nabla_b  (D_\delta
\omega_a{}^{ij})\,. \eqno(4.14)
$$
We note that Eqs.(4.6) and (4.14) are consistent with the
integrability condition, Eq.(2.12c).

This concludes the description of the deformation of
the extrinsic geometry of
the membrane.

\vskip2pc
\noindent{\bf IV.3 Deformations of worldsheet derivatives of the
 extrinsic curvature}
\vskip1pc

In theories involving terms quadratic in the extrinsic
curvatures, one needs to evaluate also
terms like $\tilde D_\delta (\tilde\nabla_a K^i_{cd})$, to
obtain the linearized equations of motion.
One would like to re-express terms of this form as,

$$\tilde\nabla_a (\tilde D_\delta K^i_{cd}) +
{\rm lower}\quad {\rm order}
\quad{\rm terms} \,,$$
and exploit the fact that we already know what $D_\delta K^i_{ab}$ is.
This involves the evaluation of the
commutator, $[\tilde D_\delta,\tilde\nabla_a ]$.
We will do this for the commutator operating on an arbitrary
worldsheet/normal frame vector, $A_{b\,i}$:

$$\eqalign{\tilde D_\delta\tilde\nabla_a A_{b\,i}
=& D_\delta\left[D_a A_{b\,i} -
\gamma_{ab}^c A_{c\,i} - \omega_{a\, i}{}^j A_{b\,j}\right]
-\gamma_i{}^j \tilde\nabla_a A_{b\,j}\cr
=& D_a D_\delta A_{b\,i} - D_\delta \left[
\gamma_{ab}^c A_{c\,i} + \omega_{a \, i}{}^j A_{b\,j} \right]
-\gamma_i{}^j \tilde\nabla_a A_{b\,j}\cr
=&D_a D_\delta A_{b\,i} -
\gamma_{ab}^c D_\delta A_{c\,i}
-\omega_{a\, i}{}^j D_\delta A_{b\,j} \cr
&\quad+ \gamma_{ab}^c D_\delta A_{c\,i}
+\omega_{a\, i}{}^j D_\delta A_{b\,j}
- D_\delta [\gamma_{ab}^c A_{c\,i}]
- D_\delta [\omega_{a\,i}{}^j  A_{b\,j}]
-\gamma_i{}^j \tilde \nabla_a A_{b\,j}
\cr
=& \tilde\nabla_a D_\delta A_{b\, i}
- ( D_\delta \gamma_{ab}^c) A_{c\,i}
- ( D_\delta \omega_{a\,i}{}^j )  A_{b\,j}
- \gamma_i{}^j \tilde \nabla_a A_{b\,j}\,. \cr
}
$$
Therefore, we find,

$$
 \left[ \tilde D_\delta , \tilde\nabla_a \right] A_{b \, i} =
- \{ (D_\delta \gamma_{ab}^c) \delta_i^j
- \left[ ( D_\delta \omega_{a\,i}{}^j )
- (\tilde\nabla_a \gamma_i{}^j) \right] \delta_b^c  \}
A_{c\, j}\,. \eqno(4.15)
$$
Note that on the right hand side appears the same covariant combination
appearing in Eq. (4.13).

A useful application of this equation is given by considering
the deformation of the d'Alembertian $\tilde\Delta =
\tilde\nabla^a \tilde\nabla_a$.
Applying the d'Alembertian to an arbitrary $\Psi^i$, one finds

$$ \eqalignno{ \tilde D_\delta ( \tilde\Delta \Psi^i )
=& ( D_\delta \gamma^{ab} ) \tilde\nabla_a \tilde\nabla_b \Psi^i
+ \gamma^{ab} \left[ \tilde D_\delta , \tilde\nabla_a \right]
\tilde\nabla_b \Psi^i
+ \gamma^{ab} \tilde\nabla_a
\{  \left[ \tilde D_\delta , \tilde \nabla_b \right] \Psi^i \}
+ \tilde\Delta (\tilde D_\delta \Psi^i ) \cr
=& \tilde\Delta (\tilde D_\delta \Psi^i )
- 2 \tilde\nabla_a \left[ K^{ab}{}_j \Phi^j
(\tilde\nabla_b \Psi^i) \right]
+  \left[ \nabla^a ( K_j \Phi^j)\right] (\tilde\nabla_a \Psi^i ) \cr
\quad +& 2 K^{ab \, [i} (\tilde\nabla_a \Phi^{k]} ) \tilde\nabla_b \Psi_k
+ 2 \tilde\nabla_a \left[ K^{ab \, [i}
(\tilde\nabla_b \Phi^{k]}) \Psi_k  \right] \cr
\quad -&  g ( R (n_j, e^b) n^k, n^i ) \Phi^j (\tilde\nabla_b \Psi_k )
- \tilde\nabla_b \left[ g ( R ( n_j, e^b ) n^k, n^i ) \Phi^j \Psi_k
\right]\,.  &(4.16)\cr
}
$$
This expression will be useful in the following section.
\vskip2pc
\noindent{\bf V. DYNAMICS: SOME EXAMPLES}
\vskip1pc

In this section, we apply the kinematical framework
we have developed  to the derivation of the equations of
motion, and of the linearized equations of motion,
for two phenomenological theories of relativistic membranes
of physical interest.
We begin with the familiar Nambu action. This will allow us to
recover the results of Refs.[3] and [4]. A second example we consider is
a correction term quadratic in the extrinsic curvatures.

The Nambu action for a relativistic membrane is given by,

$$ S_0 = -\sigma \int d^D \xi \sqrt{- \gamma }\,, \eqno(5.1)
$$
where $\sigma$ is the membrane tension.

To derive the equations of motion, we can describe the
deformations of the worldsheet with the vector field
$\delta = \Phi^i n_i$, because only motions transverse
to the worldsheet are physical.
We have that

$$ \delta S_0 = - \sigma \int  d^D \xi \sqrt{- \gamma }
\gamma^{ab} K_{ab}{}^i \Phi_i = 0\,.  $$
Therefore the equations of motion describing
an extremal surface are given by,

$$K^i = 0\,, \eqno(5.2)$$
and we recover the well known result that extremal
surfaces have vanishing trace of the extrinsic curvatures.

To obtain the linearized equations of motion, consider

$$\eqalign{\tilde D_\delta K^i =& \gamma^{ab}
\tilde D_\delta K_{ab}{}^i + K_{ab}{}^i  D_\delta \gamma^{ab}\cr
=& - \tilde\Delta \Phi^i + \left[ g(R(e_a, n_j) e^a,n^i)
+ K_{ab}{}^i K^{ab}{}_j \right] \Phi^j\,,\cr}$$
so that we find the linearized equations of motion in the form,

$$ \tilde\Delta \Phi^i + \left[
 K_{ab}{}^i K^{ab}{}_j - g(R(e_a, n_j) e^a,n^i) \right] \Phi^j = 0 \,,
\eqno(5.3) $$
in agreement with Eq.(4.1) of the second paper in Ref.[3].
This set of coupled linear equations can be seen as
the equations of motion for a multiplet of scalar fields,
with a ``variable mass" that depends on a particular
projection of the curvature of spacetime, and on the extrinsic geometry.

When the projection of the spacetime Riemann tensor vanishes,
Eq. (5.3) can be written in the form, which will be used below,

$$ {\cal O}^i{}_j \Phi^j \equiv \left[ \tilde\Delta^i_j +
 K_{ab}{}^i K^{ab}{}_j \right] \Phi^j = 0\,. \eqno(5.4)
$$

We now consider a less simple example involving an
action quadratic in the extrinsic curvature,

$$
S_2 = \alpha \int d^D \xi \sqrt{-\gamma} K_i K^i  \,. \eqno(5.5)
$$
where $\alpha$ is a coefficient characterizing the
rigidity of the membrane. This
action is of some interest in that,
modulo the totally contracted Gauss-Codazzi equation,
when $D=2$, and the background geometry is flat, this
action represents the most general action of this order
in the worldsheet geometry. For an alternative derivation of the
equations of motion corresponding to higher order actions of this
order, see Ref.[9].

The variation of this action with respect to normal
deformations of the worldsheet gives,

$$ \delta S_2 =  \alpha \int d^D \xi \sqrt{- \gamma} \left[
K_i K^i K^j \Phi_j + 2 K_i ( - \tilde\Delta \Phi^i
+ g (R(e_a,n_j)e^a, n^i) \Phi^j - K_{ab}{}^i K_{ab \,j} \Phi^j  ) \right]$$
Thus, the Euler-lagrange equations for $S_2$ are given by,

$$
\tilde\Delta K^i + \left[ - g (R(e_a,n^j)e^a, n^i) +
(\gamma^{ac} \gamma^{bd}
- {1 \over 2} \gamma^{ab} \gamma^{cd}) K_{ab}{}^j K_{cd}{}^i \right] K_j
= 0\,. \eqno(5.6)
$$
Note that extremal surfaces are obvious  non-trivial
solutions of these equations.

The linearized equations of motion are considerably more
complicated than in the case of an extremal surface.
For the sake of simplicity,
we restrict ourselves to the case in which the
background spacetime is Minkowski, in order to disregard the
spacetime curvature projections. The generalization to an
arbitrary background is  straightforward.
For this case, a lengthy computation, exploiting $(4.16)$, gives
the linearized equations of motion in the form

$$
\eqalignno{ -& \tilde\Delta \tilde \Delta \Phi^i - 2
K^{ab}{}_j K^j (\tilde\nabla_a \tilde\nabla_b \Phi^i )
+ {1\over 2} K^j  K_j \tilde\Delta \Phi^i \cr
+& ( K^i K_j - 2 K_{ab}{}^i K^{ab}{}_j ) \tilde\Delta \Phi^j
- 2 K^{ab}{}_j (\tilde\nabla_a K^j )(\tilde\nabla_b \Phi^i )
- K_j (\tilde\nabla^b K^j ) (\tilde\nabla_b \Phi^i ) \cr
 -& 2\tilde\nabla^c [ K_{ab}{}^i K^{ab}{}_j ] (\tilde\nabla_c \Phi^j )
+ 2 K^{ab\, i} (\tilde\nabla_a K_j ) (\tilde\nabla_b \Phi^j )
- 2 K^{ab}{}_j (\tilde\nabla_a K^i ) ( \tilde\nabla_b \Phi^j ) \cr
+& 2 K_j (\tilde\nabla^b K^i ) (\tilde\nabla_b \Phi^j )
- \tilde\Delta [ K_{ab}{}^i K^{ab}{}_j ] \Phi^j
- (\tilde\nabla^a K^i ) (\tilde\nabla_a K_j) \Phi^j
\cr
-& 2 K^{ab}{}_j (\tilde\nabla_a \tilde\nabla_b K^i ) \Phi^j
+ 2 K_{ab}{}^i K^{bc}{}_k K^a{}_{c\, j} K^j \Phi^k
+ {1\over 2} K_{ab}{}^i K^{ab}{}_k K^j K_j \Phi^k \cr
+& K^i K_j K_{ab}^j K^{ab}_k \Phi^k
- K_{ab}{}^i K^{ab\,j} K_{cd\,j} K^{cd}{}_k \Phi^k = 0\,. &(5.7) \cr }
$$
The scalar fields $\Phi^i$ satisfy then
a  set of coupled fourth order linear differential equations.
It is interesting to note that, for a hypersurface,
 the square of the
worldsheet d'Alembertian is the only term that depends
only on the intrinsic geometry of the worldsheet.

The linearized equations (5.7) are rather complicated. An interesting special
case is given by considering linearization about an extremal surface,
{\it i.e.} setting $K^i = 0$.
The equations simplify considerably, and reduce to
$$- \left({\cal O}^2 \right)^i{}_j \Phi^j = 0\,. \eqno(5.8)
$$
where the operator ${\cal O}$, defined in Eq. (5.4), is the
operator describing small perturbations about an extremal
surface induced by the Nambu action. It is remarkable that
its square appears here.
Thus, linear perturbations about an extremal surface
which satisfy Eq. (5.4), continue to be
solutions when one takes into account the modifications
induced by the action (5.5).

We conclude this section with the following remarks about
the deformation connection $\gamma^{ij}$.
We note that the action must be a scalar under
normal frame rotations. Such an action
(ignoring possible contractions of worldsheet or
spacetime indices) involves an integrand
consisting of a totally contracted product of normal frame tensors.
The most simple such product is of the form $P^i Q_i$,
where $P_i$ and $Q_i$ are normal frame vectors. We note that

$$\eqalign{D_\delta (P^i Q_i)
=&P^i D_\delta Q_i + Q^iD_\delta P_i\cr
=&P^i D_\delta Q_i + Q^iD_\delta P_i
-(\gamma_{ij} + \gamma_{ji}) P^i Q^j\cr
=&P^i\tilde D_\delta Q_i + Q^i\tilde D_\delta P_i
=\tilde D_\delta (P^i Q_i)\,,\cr}$$
where the second line follows from the first
line because of the anti-symmetry of $\gamma_{ij}$.
The introduction of the
normal frame covariant variation does not complicate the
derivation of the Euler-Lagrange equations. Let us now denote
these equations by,
$${\cal E}_i=0\,.$$
The perturbed equations of motion are then just
$$\tilde D_\delta {\cal E}_i =0\,.$$
Modulo the background equations of motion,
these equations reduce to $D_\delta {\cal E}_i=0$.
In other words, the connection $\gamma_{ij}$
never needs to be calculated explicitly.
In perturbation theory, the normal frame
covariant derivative comes for free.
In light of the above remarks, one can safely always set
$\gamma_{ij}=0$.

\vskip2pc
\noindent{\bf VI. DISCUSSION}
\vskip1pc

In this paper we have presented a thorough analysis of the
kinematics of infinitesimal deformations of the worldsheet spanned
by a membrane of arbitrary dimension in any
spacetime. The physical measure of the deformation is given by the normal
components of the displacement vector. These normal
components are scalar fields living on the worldsheet.
The deformation of the intrinsic geometry is straightforward.
The deformation of the extrinsic geometry, however,
is complicated by the requirement of covariance under
normal frame rotations. We introduce a
manifestly  covariant deformation operator. When we do this the
covariant deformations of both the extrinsic curvature
and the extrinsic twist curvature are given by second order hyperbolic
partial differential operators acting on the scalar fields.

This kinematical framework is applied in sec.V to derive
the equations of motion and their linearizations both for
a system described by the Nambu action and for
a system involving an action quadratic in the extrinsic curvature.
Specializing to Minkowski spacetime for simplicity, we find
that the perturbations about an extremal surface are
described by a second order hyperbolic operator
for the Nambu dynamics, and by its square for the
dynamics described by an action quadratic in the extrinsic curvature.

A more systematic treatment of all low order actions will
be addressed in a forthcoming paper [16].
We also leave for a future publication a non-perturbative description of
the deformations of a relativistic membrane.
This involves a non-trivial generalization of the
Raychaudhuri equations for a curve [17].

\vskip2pc
\centerline{\bf ACKNOWLEDGEMENTS}
\vskip1pc
We gratefully acknowledge support from
CONACyT under the grant 3354-E9308.
\vskip2pc

\centerline{\bf REFERENCES}

\vskip1pc
\item{1.} A. Vilenkin,  Phys. Rep. {\bf 121}, 263 (1985).
\vskip1pc
\item{2.} J. Garriga and A. Vilenkin,  Phys. Rev. {\bf D44}, 1007 (1991);
{\it ibid.} {\bf D45}, 3469 (1992);{\it ibid.} {\bf D47}, 3265 (1993).
\vskip1pc
\item{3.} J. Guven, {\it Phys Rev} {\bf D48} 4606 (1993);
{\it ibid.} {\bf D48} 5562 (1993); in {\it Proceedings of
SILARG VIII}, ed. by P. Letelier and W. Rodriguez,
(World Scientific, Singapore, 1994).
\vskip1pc
\item{4.} A.L. Larsen and V.P. Frolov, {\it Nucl Phys}
{\bf B414}, 129 (1994).
\vskip1pc
\item{5.} B. Carter, {\it Phys. Rev.} {D48}, 4835 (1993),
approaches the problem from a more mathematically
sophisticated point of view.
\vskip1pc
\item{6.} For example, see
H. B. Lawson {\it Lectures on Minimal Submanifolds} 2nd. edition,
(Publish or Perish, 1980).
\vskip1pc
\item{7.} For example, see  A. Polyakov, {\it Nucl. Phys.}
{\bf B268}, 406 (1986); T.L. Curtright, G.I. Ghandour, and C.K. Zachos,
{\it Phys. Rev.} {\bf D34}, 3811 (1986);
K. Maeda and N. Turok, {\it Phys Lett} {\bf B202} 376 (1988);
R. Gregory, D. Haws, and D. Garfinkle, {Phys. Rev.} {\bf D42}
343 (1990); R. Gregory, {\it ibid.} {\bf D43} 520 (1991);
B. Carter and R. Gregory, (Preprint, 1994).
\vskip1pc
\item{8.} P.S. Letelier, {\it Phys. Rev.}  {\bf D41} 1333 (1990);
B. Boisseau and P.S. Letelier, {\it ibid.} {\bf D46} 1721
(1992); See also
C. Barrabes, B. Boisseau, and M. Sakellariadou, {\it ibid.}
{\bf D49} 2734 (1994).
\vskip1pc
\item{9.}  B. Carter, {\it Class. Quantum Grav.} {\bf 11}, 2677 (1994).
\vskip1pc
\item{10.} D.H. Hartley and R.W. Tucker,
in {\it Geometry of Low-Dimensional Manifolds: 1},  London
Mathematical Society Lecture Note Series {\bf 150},  ed. by
S.K. Donaldson and C.B. Thomas (Cambridge University Press, Cambridge,
1990).
\vskip1pc
\item{11.}
L. P. Eisenhart {\it Riemannian Geometry} (Princeton Univ. Press,
Princeton, 1947);
M. Spivak {\it Introduction to Differential Geometry: Vols. I to 5 }
(Publish or Perish, Boston MA 1970);
S. Kobayashi and K. Nomizu {\it Foundations of Differential
Geometry: Volume II} (Interscience, New York 1969).
\vskip1pc
\item{12.} M. Dajczer, {\it Submanifolds and Isometric Immersions}
(Publish or Perish, Houston, Texas, 1990).
\vskip1pc
\item{13.} B. Carter, {\it Journal of Geometry and Physics}
{\bf 8}, 52 (1992).
\vskip1pc
\item{14.} B. Carter, {\it Class. Quantum Grav.} {\bf 9}, S19 (1992).
\vskip1pc
\item{15.} See, for example,
S.W. Hawking and G.F.R. Ellis {\it The Large Scale Structure of
Space-Time} (Cambridge Univ. Press, Cambridge, 1973).
\vskip1pc
\item{16.} R. Capovilla and J. Guven, (to appear).
\vskip1pc
\item{17.} R. Capovilla and J. Guven, (gr-qc/9411061).
{\it Phys. Rev.}  {\bf D} (To appear 1995).

\bye